\documentstyle[psfig]{europhys}

\newif\ifboo \boofalse

\newcommand{\be}{\begin{equation}}
\newcommand{\ee}{\end{equation}}
\newcommand{\ba}{\begin{eqnarray}}
\newcommand{\ea}{\end{eqnarray}}
\newcommand{\br}{{\bf r}}

\newcommand{\bk}{{\bf k}}

\newcommand{\bu}{{\bf u}}

\def\lesssim{\mathrel{\mathpalette\vereq<}}
\def\vereq#1#2{\lower3pt\vbox{\baselineskip1.5pt \lineskip1.5pt
\ialign{$#1\hfill##\hfil$\crcr#2\crcr\sim\crcr}}}
 
\def\gtrsim{\mathrel{\mathpalette\vereq>}}

\begin{document}

\euro{}{}{}{}
\Date{}
\shorttitle{Extension of Haff's cooling law in granular flows}

\title{Extension of Haff's cooling law in granular flows.}
\author{R.~Brito\inst{1} and M.H.~Ernst\inst{2}}
\institute{\inst{1}
Departamento  de F\'{\i}sica Aplicada I, Universidad Complutense
de Madrid \\ 28040-Madrid, Spain\\
\inst{2} Instituut voor Theoretische Fysica, Universiteit Utrecht \\
Postbus 80006,
3508 TA Utrecht, The Netherlands}

\rec{}{}

\pacs{\Pacs{05}{20.y}{Statistical mechanics}
\Pacs{05}{20.Dd}{Kinetic theory}
\Pacs{81}{05.Rm}{Porous materials; granular materials}
}

\maketitle
\begin{abstract}
The total energy $E(t)$ in a fluid of inelastic particles is 
dissipated through inelastic collisions. 
When such systems are prepared in a homogeneous initial state
and evolve undriven, $E(t)$ decays  initially as $t^{-2}
\sim \exp[ - 2\epsilon \tau]$ (known as Haff's law), where $\tau$ 
is the average number of collisions suffered by a particle within  
time $t$, and $\epsilon=1-\alpha^2$ measures the degree of inelasticity,
with $\alpha$ the coefficient of normal restitution. 
This decay law is extended for large times to $E(t) \sim 
\tau^{-d/2}$  in $d$-dimensions, far into the nonlinear clustering 
regime. The theoretical predictions are
quantitatively confirmed by computer simulations, and holds for
small to moderate inelasticities with $0.6< \alpha< 1$.
\end{abstract}

\begin{figure}[t]
\[ \psfig{file=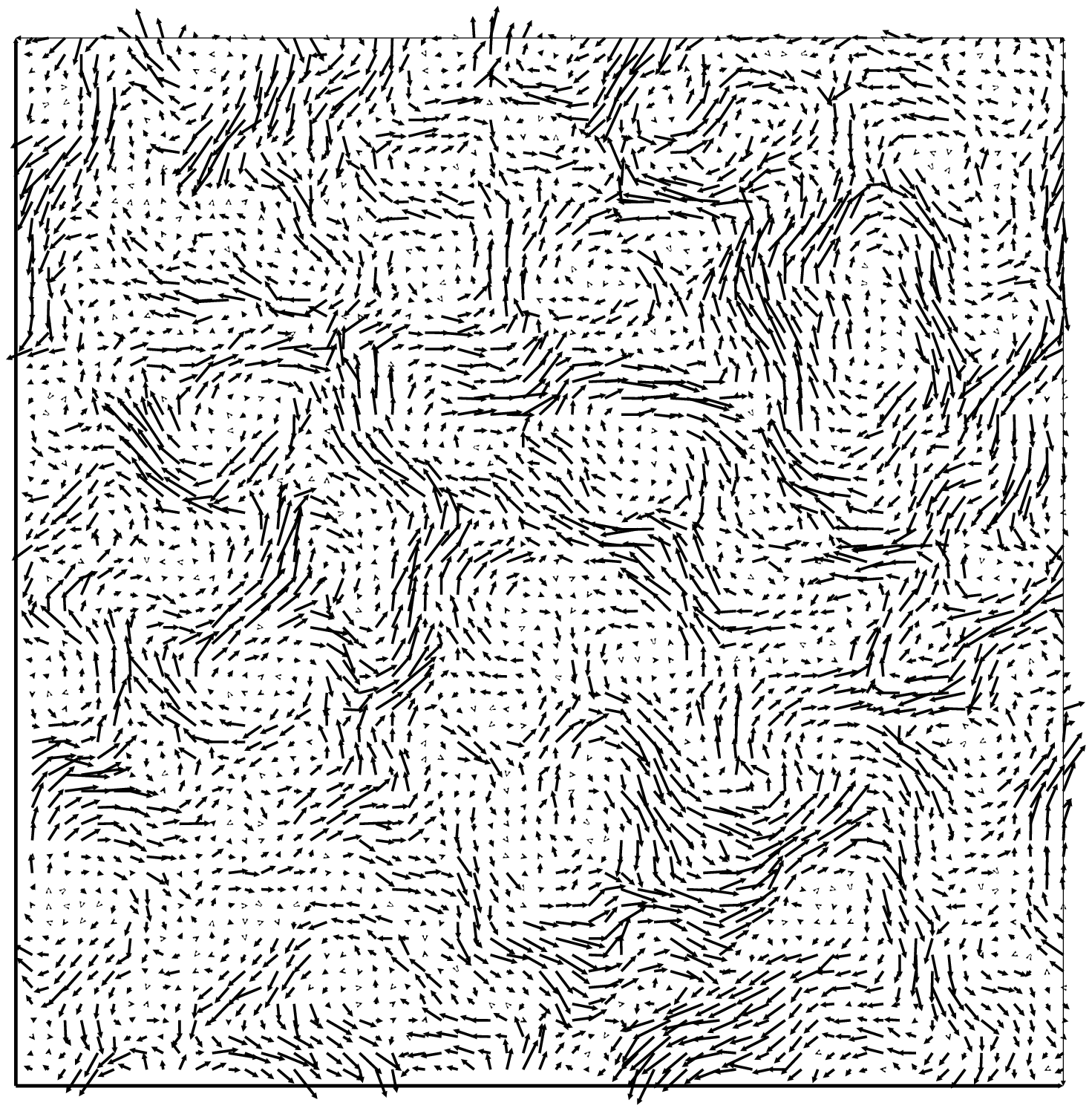,height= 7cm,angle=270} \ \ \ \
\psfig{file=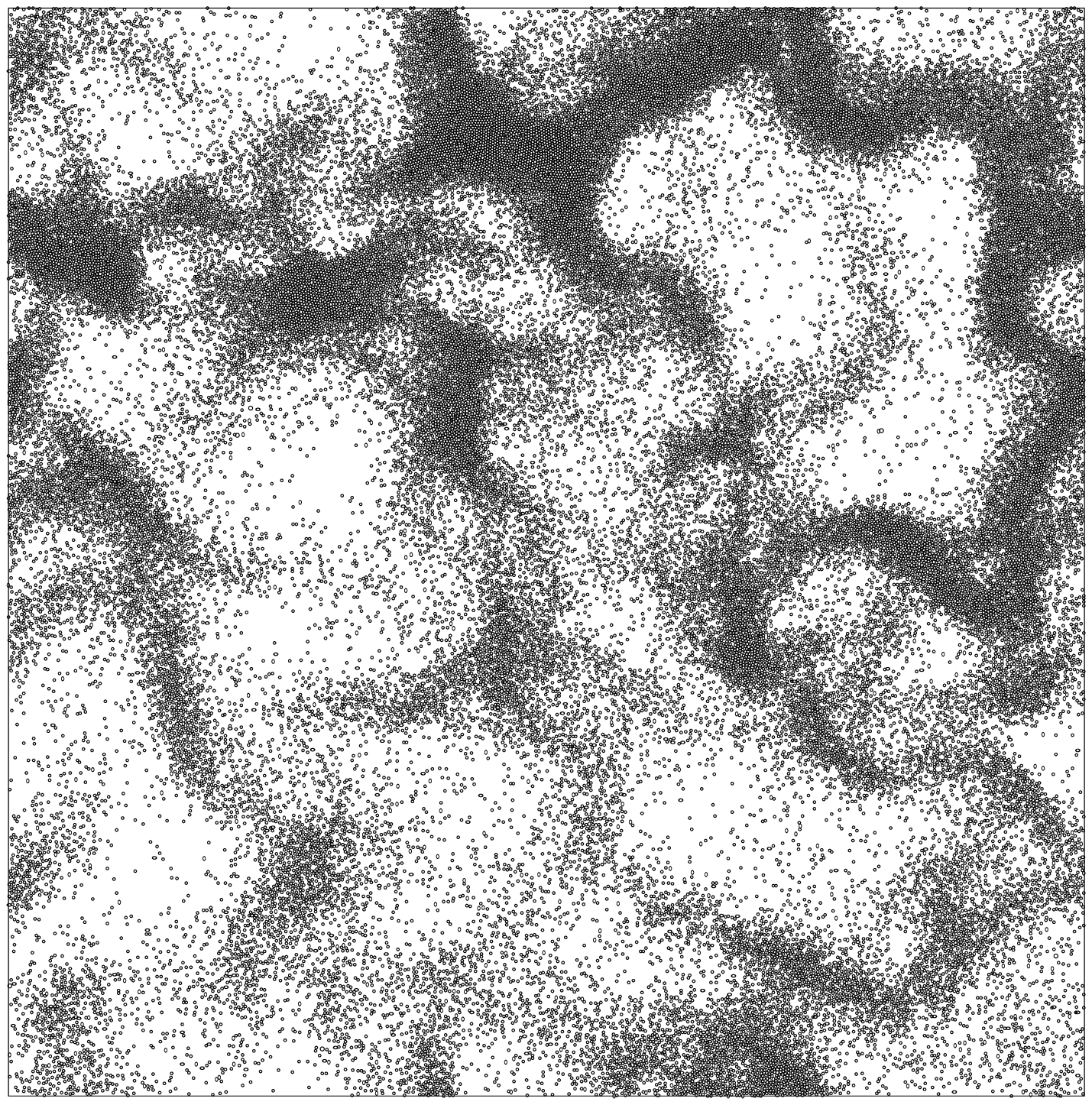,width=7cm} \]
\caption{}{Snapshots (a) of flow field $\bu(\br,\tau=80)$ and (b) of
density field $n(\br,\tau=160)$ in an IHS simulation with $N=50000,
\alpha=0.9, \phi=0.245$, prepared in an initial thermal equilibrium
state. Crossover to the nonlinear regime occurs at
$\tau_c\simeq70$. }
\label{fig1}
\end{figure}

A fluid of inelastic hard spheres (IHS) in $d$ dimensions, 
prepared initially in a state of
thermal equilibrium with temperature $T_0$, will remain for an extended
period of time (measured in average number of collisions suffered 
per particle) in a spatially homogeneous cooling state (HCS), provided 
the degree of inelasticity $\epsilon$ is small. 
In this state the average kinetic energy per particle 
$E(t)=\frac{d}{2}T(t)$ is decreasing like $1/t^2$  due 
to inelastic collisions, as first
explained by Haff \cite{Haff}. However, the HCS
with a spatially uniform density and temperature, and a vanishing 
flow field is unstable against long wavelength spatial 
fluctuations, and leads finally to a state  with large 
scale clusters in the density field  $n(\br,t)$ as well 
as in the flow field $\bu(\br,t)$ (vortices). This is
illustrated in the MD simulations of fig.~\ref{fig1}. 
In this state  cooling slows down and large deviations
from Haff's law occur, which have not yet been explained 
in any quantitative manner \cite{Goldh,US}.

In general analytic results for nonlinear decay of rapid 
granular flows are rare \cite{Goldh}.
The goal of this letter is to calculate the long time  decay of 
$E(t)$ from nonlinear hydrodynamics, using mode coupling methods. 
The basic idea is to decompose the quantities of interest, here the 
{\it total microscopic} energy, or---in the study of long time 
tails of Green-Kubo formulas---the {\it total microscopic} momentum-, 
energy- or particle fluxes, into a superposition of {\it products }
of slow hydrodynamic fluctuations of density $n(\br,t)$, temperature
$T(\br,t)$ and flow field $ \bu(\br,t)$. As one is dealing with 
fluctuations their amplitude is small in general, so their time decay
can be calculated from linear hydrodynamic modes Ref.~\cite{EHvL}, 
provided the fluctuations are stable. However, in the undriven IHS fluid 
the density fluctuations are growing exponentially like 
$\exp(\epsilon\tau)$ (clustering instability) \cite{Goldh}.
The redeeming feature in the resolution of this paradox is 
that the energy only depends on products of modes, in 
which exponentially growing density fluctuations are 
balanced by exponentially decreasing temperature fluctuations.

In the case of {\it inelastic} hard spheres two
colliding  particles loose a fraction, $\epsilon\equiv 1-\alpha^2$, of
their relative kinetic energy, where  $\epsilon$ is the degree of
inelasticity and $\alpha$ the coefficient of normal restitution.
Particles move more and more parallel after every collision,
and their random motion (temperature) decreases;
{\em i.e.} the spatial fluctuations at shorter wavelength (which 
are rapidly damped out through viscosity and heat conduction)
are not replenished with energy from the more microscopic degrees 
of freedom, as in the elastic case, where fluctuations at all 
wavelengths are kept at thermal noise level, fed by the randomizing
elastic collisions.
So, the dynamics selectively suppresses the shorter wave length components
of the flow field, as is clearly illustrated in fig.~1a. 
After a time on the order of the mean free time $t_0$ between
collisions, the surviving fraction of the energy $E(t)=(1/N)
\langle\sum_i\frac{1}{2} mv_i^2\rangle$ is totally stored in hydrodynamic
modes (with wavenumbers $k<1$, where $k$ is the measured in units of inverse
mean free path $l_0$) of the kinetic energy of the flow and the
internal energy (granular temperature),
\be
E(t)\simeq \frac{1}{N} \int \mbox{d}\br \left\{ \frac{1}{2} 
\langle \rho(\br,t) u^2(\br,t)\rangle +\frac{d}{2} \langle 
n(\br,t) T(\br,t)\rangle\right\} .
\label{Et}
\ee
The subsequent decay of $E(t)$ is controlled by two different
$\bk$-regimes \cite{McNam}: (i) the {\em elastic} regime
($\epsilon < k< 1$) at short times (homogeneous cooling regime)
with $t_0<t<t_0/\epsilon$, and (ii) the {\em dissipative} regime
($k<\epsilon$) at large times ($t\gg t_0/\epsilon$).

In the first regime the decay is controlled by the HCS, where 
$n(\br,t)=n$, $\bu(\br,t)=0$ and a homogeneous temperature, 
decaying according to Haff's cooling law, 
\be
T(t)=\frac{T_0}{(1+\gamma_0 t/t_0)^2} \equiv  
T_0 \exp (-2\gamma_0 \tau),
\label{HCS} \ee
where $E_0=\frac{d}{2} T_0$ is the initial (equilibrium) 
energy per particle, $\gamma_0=\epsilon/2d$, and 
$t_0= 1/\omega(T_0)$ is the mean free time with 
collision frequency $\omega(T_0)$ in the initial state. 
The second equality in eq.~(\ref{HCS}) defines the average number
of collisions per particle within a time $t$. It is
obtained by integrating $d\tau=\omega(T(t)) dt$, 
where $\omega(T)\sim \sqrt{T}/l_0$ with a temperature independent 
mean free path $l_0$, given by the Enskog theory for a 
dense system of hard spheres ($d=2,3$). Haff's law is illustrated 
in figs.~2a,b by the straight lines, labeled 1. The HCS is 
essentially an {\em adiabatically} changing equilibrium state, 
parametrized by $\{ n,\bu=0,T(t)\}$. 

To evaluate the contribution of the long wavelength fluctuations, we 
linearize the fields in eq.~(\ref{Et}) around the HCS, where
$\bu(\br,t)$, $\delta n(\br,t)=n(\br,t)-n$ and 
$\delta T(\br,t)=T(\br,t)-T(t)$, are small (at least 
their combined effects; see redeeming features below), and
consider Fourier modes. Retaining up to quadratic terms  yields, 
\be
E(t)= \frac{d}{2} T(t) +\frac{1}{2N} \int \frac{\mbox{d}\bk} {(2\pi
l_0)^d} \left\{ \rho \langle |\bu(\bk,\tau)|^2\rangle + d\langle\delta
n(\bk,\tau) \delta T (-\bk,\tau\rangle\right\},
\label{Et2}\ee
where $T(t)$ is given by Haff's law. The first term in the integrand 
is proportional to the structure factors 
$[(d-1)S_\perp (k,\tau) +S_\parallel(k,\tau)]$, and the second 
one to $S_{nT}(k,\tau)$, where the subscript $\parallel$ 
refers to the longitudinal component of $\bu$ and 
the subscript $\perp$ to the transverse ones.

According to the theory of ref.~[3a,b], the fluctuations 
around the HCS are described by fluctuating hydrodynamic 
Langevin equations, and obey an `adiabatic' fluctuation 
dissipation theorem. The results from the structure factors,
$S_{ij}(\bk, \tau)$, obtained by numerically integrating
the fluctuating hydrodynamic equations of ref. [3b], could be  
integrated over the full hydrodynamic range of wave numbers to 
obtain the decay of $E(t)$ for all times $t>t_0$. 

Here, however, we present an analytic method that enables 
us to calculate the asymptotic decay for $t\gg t_0/\epsilon$
explicitly. For such times only modes in the dissipative regime with 
$k\ll\epsilon$ will contribute to the integrands in (\ref{Et2}). 
In this regime, the dispersion relations for the modes, 
relative to the HCS, are well known [5,3c].

The structure factors for the $(d-1)$ transverse components 
$u_{\perp\alpha}$ and the longitudinal one $u_l$ in the range 
$k\ll \gamma_0\sim\epsilon $ can be deduced from ref.~[3b],
and yield, 
\be
V^{-1} \langle| u_{\perp\alpha}(k,\tau)|^2\rangle=(T_0/\rho) 
e^{-2\Delta_\perp k^2\tau} \left( 1+\Delta_\perp k^2/\gamma_0\right),
\label{uperp} 
\ee
with a diffusivity $\Delta _\perp =\nu /l_0^2 \omega$, 
where $\nu=\eta/\rho$ is the kinematic viscosity. The longitudinal
structure factor is given by a similar expression with $\Delta _\perp$
replaced by $\Delta _\parallel$. The explicit form of the
longitudinal diffusivity (given in ref.~[3b] as 
$\Delta _\parallel=\gamma_0\xi_\parallel^2/l_0^2$) 
is not needed here. One can verify from the explicit expressions
that $\Delta _\parallel > \Delta _ \perp$, and even 
$\Delta _\parallel \gg \Delta _ \perp$ for $\gamma_0\to0$. 

It should be noted that the dimensionless diffusivities, 
$\Delta_\perp$ and $\Delta _\parallel $, do not depend on the
local temperature, since transport coefficients and collision
frequencies are proportional to $\sqrt T$. 

By inserting eq.~(\ref{uperp}) into (\ref{Et2}) and performing 
the $\bk$-integrals, one easily finds that the contributions 
$E_{uu}(t)$ of the term $\langle |u|^2\rangle$ to eq.~(\ref{Et2})
decay as,
\be 
E_{uu}(t)\simeq \frac{T_0}{2n l_0^d} \left\{\frac{d-1}{(\Delta_\perp)^{d/2}} +
\frac{1}{(\Delta_\parallel)^{d/2}} \right\}\left( \frac{1}{8\pi
\tau}\right)^{d/2}.
\label{Euu} \ee
Next, we consider the contributions $E_{nT}(t)$ in (\ref{Et2})
from the $n$-$T$ fluctuations. Calculation of the time dependence 
of $\delta n(\bk,\tau)$ and $\delta T(\bk,\tau)$ in the 
dissipative regime is in principle straightforward, but technically 
much more involved. One solves the initial value problem for the 
coupled linearized hydrodynamic equations by determining its
eigenvalues (to ${\cal{O}}(k^2)$-terms included), and eigenvectors
(to ${\cal{O}}(k)$-terms). The result for $k\ll \epsilon$ is, 
\ba
\delta n(\bk,\tau) & \simeq& \frac{ik}{\gamma_0} C_1 
e^{(\gamma_0 -\Delta_\parallel k^2)\tau} u_l(\bk,0)+\cdots 
\nonumber \\
\delta T(\bk,\tau) & \simeq& \frac{ik}{\gamma_0} C_2 
e^{-(\gamma_0 +\Delta_\parallel k^2)\tau} u_l(\bk,0)+\cdots ,
\label{deltas} \ea
where the dots represent subdominant decaying terms, and $C_n
(n=1,2)$ are some constants. The technical details are 
available through ref.~\cite{cond-mat}. 

The first equation in (\ref{deltas}) exhibits the 
{\em clustering instability}. It shows that the density
fluctuation is growing at an exponential rate. We are not
aware of any publication on a freely evolving rapid granular 
flows or IHS fluids, that explicitly demonstrate how the 
clustering instability is driven through the coupling to 
the long wavelength longitudinal velocity fluctuations---except for
the closely related nonlinear slaving mechanism between density and flow 
field, discussed by Goldhirsch {\em et al.} \cite{Goldh}---, whereas 
all other fluctuations remain bounded by their values in the initial
thermal equilibrium state, {\em i.e.} the fluctuations in the flow field
{\em decay diffusively} at rates $\Delta_\perp k^2$ and 
 $\Delta_\parallel k^2$, and those in the temperature 
decay rapidly at a rate $(\gamma_0 +\Delta_\parallel k^2)$. 

The above discussion also implies that an {\em incompressible} 
granular flow ($\nabla\cdot\bu=0$) 
would not exhibit a clustering instability, at least 
not in a linear stability analysis. Moreover, an initial 
density fluctuation $\delta n(\bk,0)$ would not cause any clustering 
instability.  Furthermore, we observe 
that $\langle \delta n(\bk,\tau) \delta T(-\bk,\tau)\rangle $
remains bounded, because the exponential increase of the first factor
is balanced  by an exponential decrease of the second.
The resulting integral yields $ E_{nT}(t) \propto  
{\cal{O}}(\tau^{-d/2-1})$, which is a
subleading asymptotic term when compared to (\ref{Euu}).

It can be shown similarly that the three mode 
contribution $E_{nuu}(t)$, 
neglected in the transition from eq.~(\ref{Et}) to (\ref{Et2})
remains bounded, and decays as $\tau^{-d}$, again a subleading 
correction. The dominant large time contribution $E_{uu}(t)$ is plotted 
in figs.~2a,b, where the lines labeled 2 and 3 represent the 
contributions of the transverse and longitudinal modes,
respectively. The former contribution 
is a factor $e^3\simeq 20$ larger than the latter.  

The result  (\ref{deltas}) also shows that there is no balancing 
of unstable density fluctuations in the structure factor 
$S_{nn}(k,\tau)\sim \langle |\delta n(\bk,\tau)|^2\rangle$. 
Consequently, the predictions of the hydrodynamic Langevin 
equations for $S_{nn}$ breakdown at the crossover from the
homogeneous cooling regime, as explained in ref.~[3b], whereas 
those for $S_\perp$, $S_\parallel$ and $S_{nT}$, used in this paper
are expected to remain valid until far into the nonlinear clustering
regime.  
 
In summary, in the decay of the total energy $E(t)$, one can distinguish
two different behaviors:  homogeneous cooling for $\tau< \tau_c$, where
$E_{\mbox{\scriptsize HCS}}(t)\simeq \frac{d}{2} T_0 
\exp(-2\gamma_0 \tau)$ is given by 
Haff's law, and a  diffusive decay for $\tau> \tau_c$, where
$E(t) \simeq E_{uu}(\tau)$ is solely determined by the 
flow field, {\it all} \,long wavelength components of which 
decay diffusively. The equality
$E_{uu} = E_{\mbox{\scriptsize HCS}}$ defines the crossover 
time $\tau_c$, typical values
of which are listed in the figure captions and table~1.
 
Before comparing the preceding results with MD simulations, 
a final comment about the relation between the `internal' kinetic time $\tau$
and the `external' time $t$ is needed: 
in the homogeneous cooling state the relation between both
times is correctly predicted by $\tau=(1/\gamma_0) \ln (1+\gamma_0t/t_0)$
in eq.~(\ref{HCS}). 
In the nonlinear clustering regime the relation between
both times is not understood. The value of $\tau (t)$, as measured in 
the computer simulations for $\tau > \tau_c$, increases 
{\it faster} than those given by (2).
  For the cases corresponding to figs.~\ref{fig1}a,b the 
time $\tau$ becomes roughly linear in $t$ for $\tau$
larger than 80.

\begin{figure}[h]
\[ \psfig{file=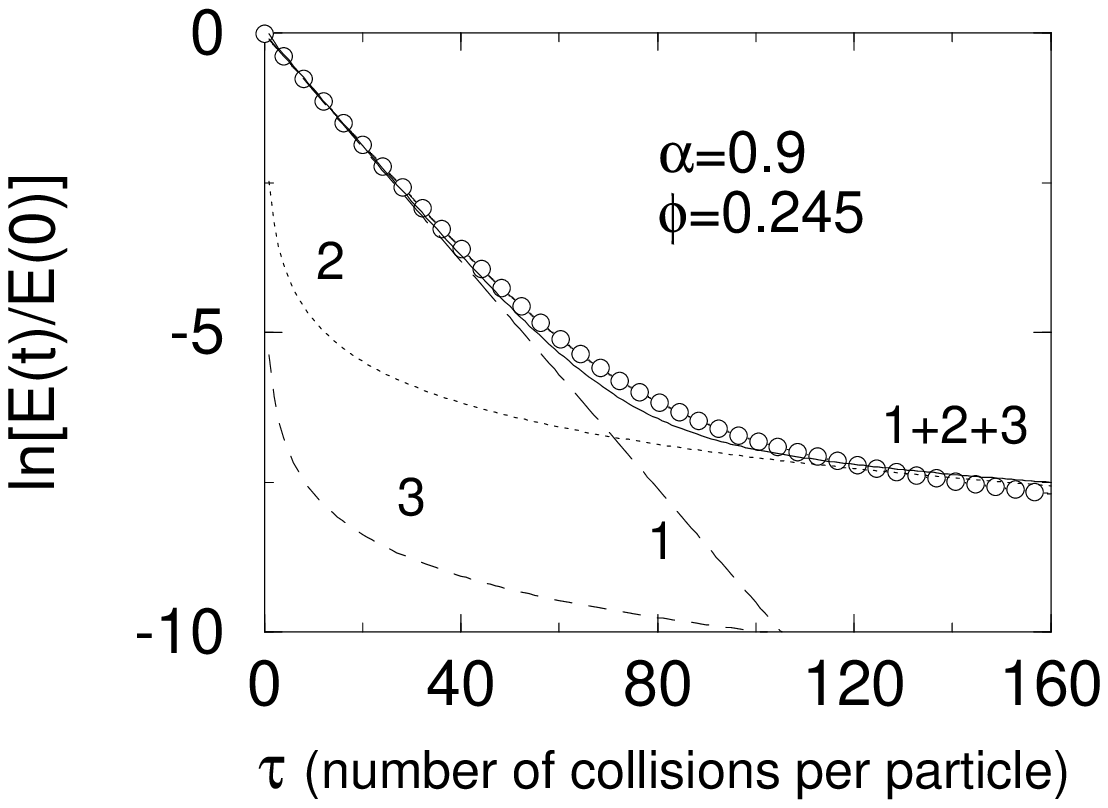,width=7cm,angle=270} \ \ \
\psfig{file=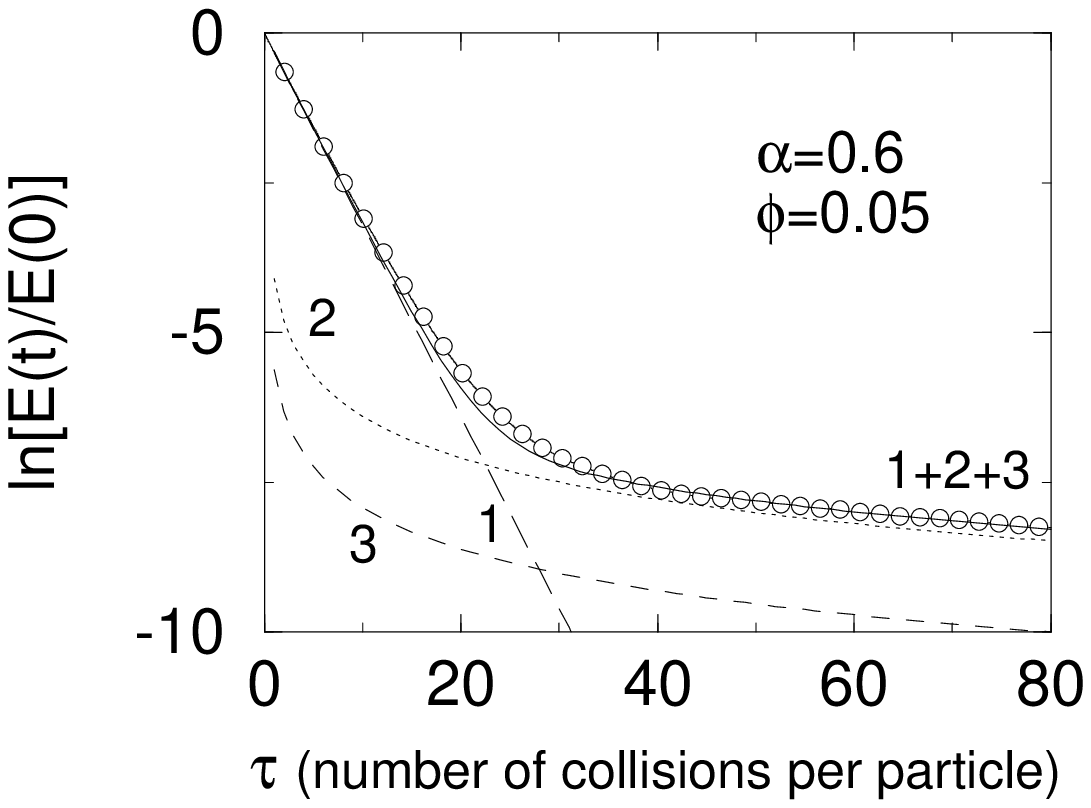,width=7cm,angle=270} \]
\caption{}{Energy decay $E(t)$~{\em vs.}~$\tau$ in simulations ($\circ$),
with $N=50000$ IHS, (a) at low inelasticity
($\alpha=0.9, \phi=0.245, \tau_c\simeq 70$), and (b) at high
inelasticity ($\alpha=0.6, \phi=0.05, \tau_c\simeq23$), compared
with theoretical prediction (labeled as 1+2+3, solid line), containing
homogeneous cooling (1), vorticity diffusion (2) and longitudinal
diffusion (3). At the latest simulation times ($\tau_l$ in Table 1) both
systems are in the nonlinear clustering regime (see fig.~\ref{fig1}b).}
\label{fig2}
\end{figure}

In fig.~\ref{fig2}a the theoretical prediction for $E(t)$ is
compared with computer simulations in a dense system $(\phi =0.245)$
of $N=50000$ IHS particles at {\it low} dissipation $(\alpha=0.9)$.
The density or coverage is defined as $\phi=\frac{1}{4} \pi N \sigma^2/L^2$, 
where $\sigma$ is the diameter of a disk. The agreement	remains good until 
far into the nonlinear  clustering regime with $\tau\gg\tau_c\simeq 70$
(see fig.~\ref{fig1}a,b).  The density 
field $n(\br,t)$ contains small ($\tau=80$) and large ($\tau=160$; see
fig.~\ref{fig1}b) spatial inhomogeneities. The flow field shows both
at $\tau=80$ (see fig.~\ref{fig1}a) and at $\tau=160$ well-developed
vortex patterns.  To describe the crossover regime around $\tau_c$ one
would have to evaluate $E(t)$ in the total 
hydrodynamic time regime ($t>t_0$), by performing the numerical calculations
described in the paragraph below eq.~(\ref{Et2}). 
The same good agreement is found at small inelasticities 
($\alpha=0.975, 0.9$) for all densities ($\phi=0.05, 0.11,  0.245,0.4$) 
in the larger systems with $N=50000$ particles, as summarized in Table 1. 
When the number of particles is reduced to $N=20000$ and $N=5000$, 
the deviations steadily increase, suggesting finite size effects.

Figure \ref{fig2}b shows a similar plot of $E(t)$ in a low density system
($\phi=0.05$) with a rather large inelasticity ($\alpha = 0.6$), 
and consequently a short crossover time $\tau_c=23$. 
There is very good agreement between theory and
simulations over the whole time interval, 
until the time of the latest measurement $\tau_l=80$. At still 
larger inelasticities and higher densities ($\alpha=0.4, \phi=0.4$)
the theoretical prediction is an 
order of magnitude smaller than the simulated $E(t)$.

\begin{table}[t]\begin{center}
\begin{tabular}{|l|l|r|r|r|c|} \hline
\multicolumn{1}{|c}{$\alpha$}&
\multicolumn{1}{|c}{$\phi$}&
\multicolumn{1}{|c}{$\tau_c$}&
\multicolumn{1}{|c}{$\tau_\parallel$}&
\multicolumn{1}{|c}{$\tau_l$}&
\multicolumn{1}{|c|}{$E_{sim}/E_{theor}$} \\ \hline
0.975  & 0.245  & 340  &27 & 500  &0.65$^*$ \\ \hline
0.9    & 0.245  &70  & 240  & 160  & 0.8  \\  \hline
0.9    & 0.4    &  68  & 56 &  160  &   1   \\ \hline
0.6    & 0.05   &  23  &$10^4$&  80   & 1    \\ \hline
0.6    & 0.11   &  19  &3000 & 160  & 1.4 \\  \hline
0.6    & 0.4    &  15  &150  & 80   & 4   \\ \hline
0.4    & 0.4    &  11  &190  & 80   & 10   \\ \hline
\end{tabular} \end{center}
\caption{Comparison $E_{sim}(\tau_l)$ with $E_{theor}(\tau_l)$
(\/$^*$$N=2\times 10^4$; in all others cases  $N=5\times 10^4$).}
\end{table}

Table 1 also shows that the deviations at $\alpha=0.6$ between theory
and simulations increase with increasing density 
(see $\phi=0.05, 0.11, 0.4$), suggesting that the increase of 
$\Delta (m,\alpha)/\Delta (m,1)$ with $ m = (\parallel, \perp)$ as
$\alpha\to 0$ is much larger at high than at low density. 
A possible explanation might be that the Enskog theory for a 
dense IHS fluid, which is also based on molecular chaos, breaks down 
at higher inelasticities. The effects of dynamic correlations are 
expected to increase strongly  with decreasing $\alpha$, 
when frequent inelastic collisions
force the particles to move more and more parallel.
But so far ring kinetic theory \cite{kin} for calculating 
transport coefficients in IHS fluids has not yet been developed.

The present theory is only applicable to thermodynamically large
systems, because in deriving (\ref{Et2}) the $\bk$-sums have been
replaced by $\bk$-integrals. In the limit of small inelasticities
($\gamma_0\to0$), the smallest wave number is $k\sim 1/L$,
and the restriction $k\ll\gamma_0$ might be violated at small $L$.

For finite systems deviations from these predictions may occur because
of interference effects through the periodic boundaries. In elastic hard
sphere fluids such effects can be expected for times larger than the
acoustic traversal time of the system. Here the fastest spreading mode
is the longitudinal diffusion, and one may expect interference effects
when the relevant diffusion length, $\xi_\parallel \equiv
\sqrt{8\Delta_\parallel \tau} $, satisfies $\xi_\parallel \gtrsim
\frac{1}{2}L$. This implies $\tau \gtrsim \tau_\parallel \equiv
L^2/32\Delta_\parallel$.  For the simulations in fig.~\ref{fig1} and
\ref{fig2}a the time of the latest measurements, $\tau_l =160$, 
is of the order of 
$\tau_\parallel =240$, and the ratio $R(\tau_l)\equiv E_{sim}/E_{theor}
\simeq 0.8$. In fig.~\ref{fig2}b, $\tau_l=80\ll\tau_\parallel=10^4$, 
and $\tau_\parallel =23$, and the ratio $R(\tau_l)=1$.

In summary, the mode coupling theory, applied to a freely evolving
IHS fluid at small inelasticity $\epsilon$, explains how the 
hydrodynamic modes in the {\em elastic} range ($\epsilon \ll k < 1$) 
are responsible for Haff's homogeneous cooling law (\ref{HCS}) 
at times $t\lesssim t_0/\epsilon$, and those in the 
{\em dissipative} range ($k\ll \epsilon$) for the 
{\em diffusive decay}, $E(t)\sim \tau^{-d/2}$ valid for $t\ll t_0/\epsilon$. 
In fact, the full range of hydrodynamic times ($t>t_0$) can be 
evaluated numerically by applying the theory of refs.~[3a,b]
over the full range of hydrodynamic wavenumbers ($k<1$).
The solid line in figs.~\ref{fig2}a,b, representing the sum of homogeneous
cooling and transverse and longitudinal velocity diffusion, is simply
an interpolation between the theoretical long and short 
time results. The agreement between theory and simulations is
excellent over the large time interval from homogeneous
cooling until far into the nonlinear clustering regime.

We also observe that the motion of the IHS fluid at the
largest $\tau$-values (see fig.~1a,b and 2a,b) is 
almost `frozen in', as the kinetic energy has typically decayed by a factor
of order $10^{-3}$. Therefore, we do not expect a later crossover
to truly nonlinear regime, which would imply that the 
decay law, derived here, is asymptotic indeed. 
At {\em large} inelasticities ($\alpha\lesssim 0.4$) the
present theory, which is based on a separation of time scales for 
small $\epsilon$, is not applicable, and the value of $E(t)$, 
observed in the simulations, is an order of magnitude larger than the 
prediction of the present theory.

Acknowledgements: It is a pleasure to thank D.~Frenkel for an invaluable
comment and helpful
correspondence. We thank T.P.C. van Noije, J.A.G.~Orza,
I.~Pagonabarraga  and M.~Hagen for stimulating discussions.
The authors also acknowledge financial
support from the Offices of International
Relations of Universidad Complutense and Universiteit Utrecht.
One of us (R.B.) acknowledges support to DGICYT (Spain) number PB94-0265.

\end{document}